\documentclass{article}
\usepackage{spconf,amsmath,graphicx}
\usepackage{bbding}
\usepackage{epsfig,amssymb,amsmath}
\usepackage{amsmath,graphicx}
\usepackage{array, amsfonts}
\usepackage{booktabs}
\usepackage{makecell}
\usepackage{color}
\usepackage{multirow}
\usepackage{bm}
\usepackage{newtxtext,newtxmath}

\title{An attention-based backend allowing efficient fine-tuning of transformer models for speaker verification}

\name{Junyi Peng$^{1}$, Old\v{r}ich Plchot$^{1}$, Themos Stafylakis$^{2}$,  Ladislav Mo\v{s}ner$^{1}$, Luká\v{s} Burget$^{1}$, Jan \v{C}ernocký$^{1}$}
\address{
$^1$Brno University of Technology, Faculty of Information Technology, Speech@FIT, Czechia \\
$^2$Omilia - Conversational Intelligence, Athens, Greece
}

\begin{document}
\ninept
\maketitle
\begin{abstract}
In recent years, self-supervised learning paradigm has received extensive attention due to its great success in various down-stream tasks. However, the fine-tuning strategies for adapting those pre-trained models to speaker verification task have yet to be fully explored. In this paper, we analyze several feature extraction approaches built on top of a pre-trained model, as well as regularization and a learning rate scheduler to stabilize the fine-tuning process and further boost performance: multi-head factorized attentive pooling is proposed to factorize the comparison of speaker representations into multiple phonetic clusters. We regularize towards the parameters of the pre-trained model and we set different learning rates for each layer of the pre-trained model during fine-tuning.
The experimental results show our method can significantly shorten the training time to 4 hours and achieve SOTA performance: 0.59\%, 0.79\% and 1.77\% EER on Vox1-O, Vox1-E and Vox1-H, respectively.
\footnote{The code will be available with the submission of the final paper.}


\end{abstract}
\section{Introduction}
\label{sec:intro}

Speaker verification (SV) is a process to verify whether an unknown speech belongs to its claimed identity.  
Since deep learning has shown superior results in speech processing, intensive research focuses on building deep structures \cite{snyder2018x,desplanques2020ecapa} to extract discriminant speaker representations. These speaker embedding extractors are typically based on CNNs and trained from scratch on a fully supervised dataset. Given the substantial size of these models and sometimes limited training data for a specific domain, it can be challenging to properly train them to capture important speaker-specific features and achieve good robustness across domains.


\begin{figure}[htb]
  \centering
  \includegraphics[width=\linewidth]{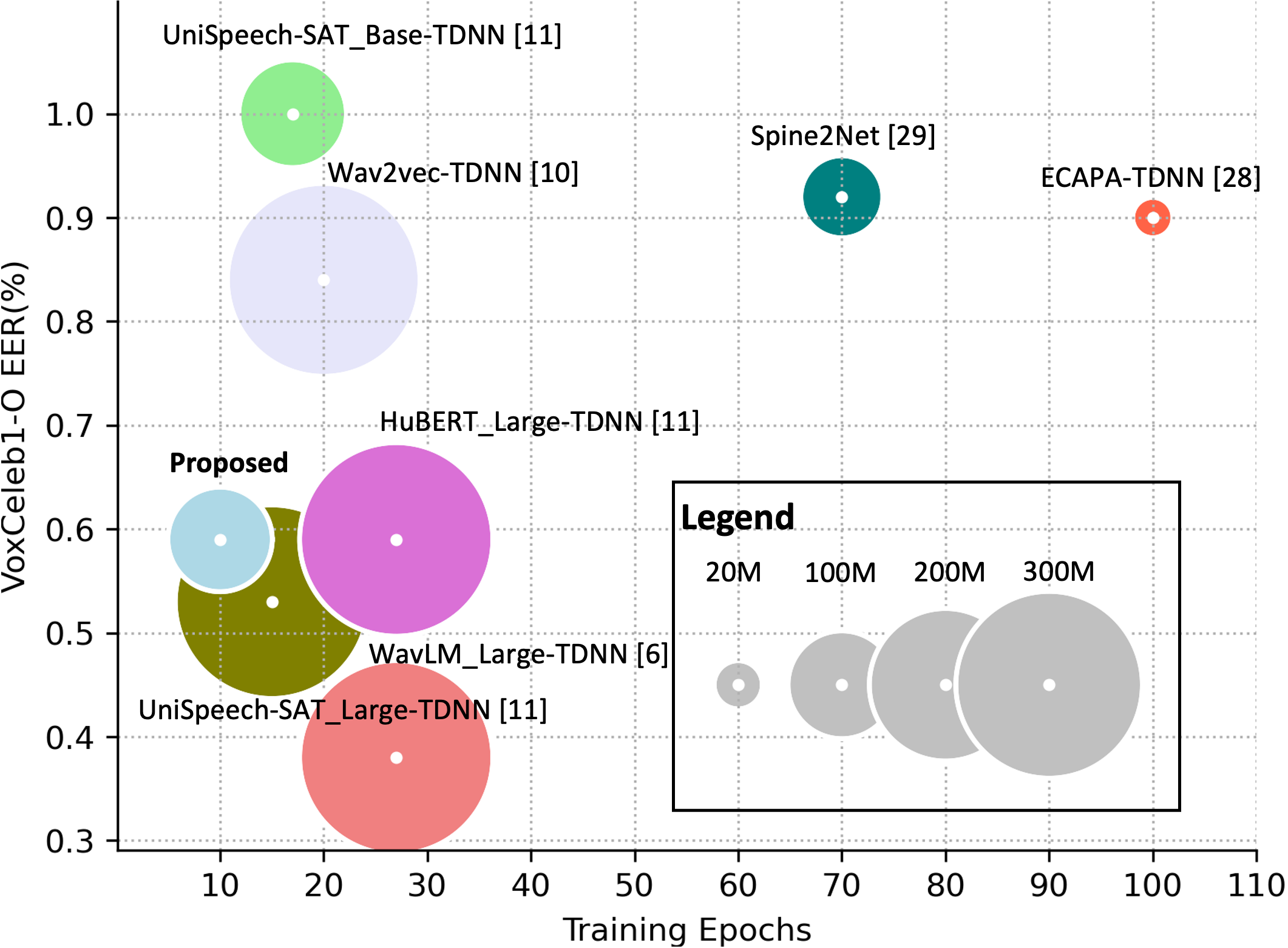}
\caption{ \textbf{EER vs training iterations, size $\propto$ learnable parameters.} Equal Error Rate (EER) on VoxCeleb1-O versus reported training epochs. The area of each circle is proportional to the total number of speaker extractor learnable parameters, while the centers of circles represent EER.} 
\label{fig:eer_epoch}
\end{figure}

Recently, the large (\textasciitilde 100M parameters and more) pre-trained Transformer models, including Wav2Vec \cite{schneider2019Wav2Vec, baevski2020Wav2Vec}, HuBERT \cite{hsu2021hubert}, WavLM \cite{chen2021wavlm}, and their variants \cite{chen2022unispeech} have significantly boosted the performance in the field of speech processing. 
These models are optimized in an unsupervised way for a speech prediction task on large-scale unlabeled data, which urges them to model the structure of speech via learning various acoustic units (such as phonemes, syllables, or other lower-level representations). By being trained on large corpora (e.g. 94K hours), these models provide an excellent initial feature extractor for a down-stream speech tasks, which can lead to faster training\cite{yang2021superb}. 

The performance of a task at hand can be further boosted through their joint fine-tuning with a task-oriented back-end (classifier, embedding extractor, etc.) For example, in \cite{fan2020exploring}, an average pooling layer and a fully connected layer are added on top of the Wav2Vec encoder, and then jointly fine-tuned in a multi-task framework. 
In order to further improve the speaker verification performance, in \cite{chen2022unispeech, novoselov2022robust}, a more powerful back-end is employed, which includes time-delay neural network (TDNN) \cite{desplanques2020ecapa} and statistical pooling layer. These methods achieve excellent performance, but the fairly large back-end size and its architecture makes it vulnerable to overfitting and complicates the training.
To alleviate this problem, in \cite{chen2022large,chen2021wavlm}, a two-stage fine-tuning strategy is investigated: at first, the pre-trained parameters are frozen, and the randomly initialized speaker embedding extractor back-end is optimized from scratch, then, both models are trained jointly. This strategy leads to more stable training process and better results obtained on the VoxCeleb data-set, but it is computationally expensive as one needs to spend many iterations propagating through the fixed pre-trained model only to initialize the back-end and only then the full power of the architecture can be exploited via the joint training.

In this paper, we investigate how to maximize the power of pre-trained  models in speaker verification task by introducing a light-weight back-end and designing proper fine-tuning strategies. Our back-end is solely based on attention and contains \emph{no convolutional layers} (neither 1D as in TDNNs, nor 2D as in ResNets). The time position of each frame is encoded only through the inherent mechanism of each Transformer model (e.g. positional embeddings). As shown in Fig \ref{fig:eer_epoch}, the proposed method demonstrates competitive SV performance as well as more efficient training process over existing state-of-the-art pre-trained model based SV systems that are integrated with TDNN. 
The contributions of our work are as follow:
\begin{itemize}
    \item We propose a light-weight, convolution-free back-end model consisting of multi-head factorized attentive pooling (MHFA) and a linear layer to extract speaker representations. As demonstrated by \cite{yang21c_interspeech}, lower layers of pre-trained models are more effective for speaker verification while top layers focus on phonetic information. The proposed method exploits all levels of representations in the pre-trained model. It focuses not only to low-level spectral characteristics carrying most of the speaker information, but it also takes into account phonetic content of the utterances.
    
    \item We explore several strategies to stabilize the fine-tuning process and to boost the performance. We experiment with assigning each layer a different learning rate and for the pre-trained model fine-tuning, we employ L2 regularization towards its original parameters.
 \item We demonstrate the effectiveness of the proposed light-weight back-end, as well as the fine-tuning strategies, when integrating with three popular pre-trained models: Wav2Vec 2.0, HuBERT, and WavLM. 
    \item We achieve state-of-the-art results on VoxCeleb data-set using fewer training iterations using a small attention-based back-end. With the same pre-trained model, the proposed system outperforms the WavLM-TDNN \cite{chen2021wavlm} and yields 0.59\%  EER on Vox1-O. 
    In addition, compared to WavLM-TDNN, the proposed method can significantly shorten the training time from 30+ hours to only 4 hours on the same computing node. 
\end{itemize}

The rest of the paper is organized as follows. Section~\ref{sec:PW} presents methods for extracting speaker information from pre-trained transformer models and the proposed method. Section~\ref{sect:over-fit} suggests two methods for avoiding over-fitting when fine-tuning. Section~\ref{sec:exp} describes the experimental setup and analyzes the results. Section \ref{sec:conclusion} concludes the paper.



\section{Extracting speaker information from pre-trained models}
\label{sec:PW}

Most of the works that attain state-of-the-art performance in speaker recognition using Transformed-based pre-trained models are utilizing them as feature extractors, and feed their features to a TDNN or ECAPA-TDNN \cite{fan2020exploring,novoselov2022robust,chen2021wavlm}. Other approaches~\cite{wang2021fine,vaessen2022fine} fine-tune the pre-trained model using only very lightweight back-end (i.e directly pool the features from the pre-trained model to extract embeddings, which are then fed to a simple linear classifier), but yield inferior performance compared to the former. The main goal of this paper is to explore methods that do not require deep and convolutional architectures and preserve the attention-based nature of the pre-trained 
network, while attaining state-of-the-art performance.

\begin{figure}
    \centering
    \includegraphics[height=5cm]{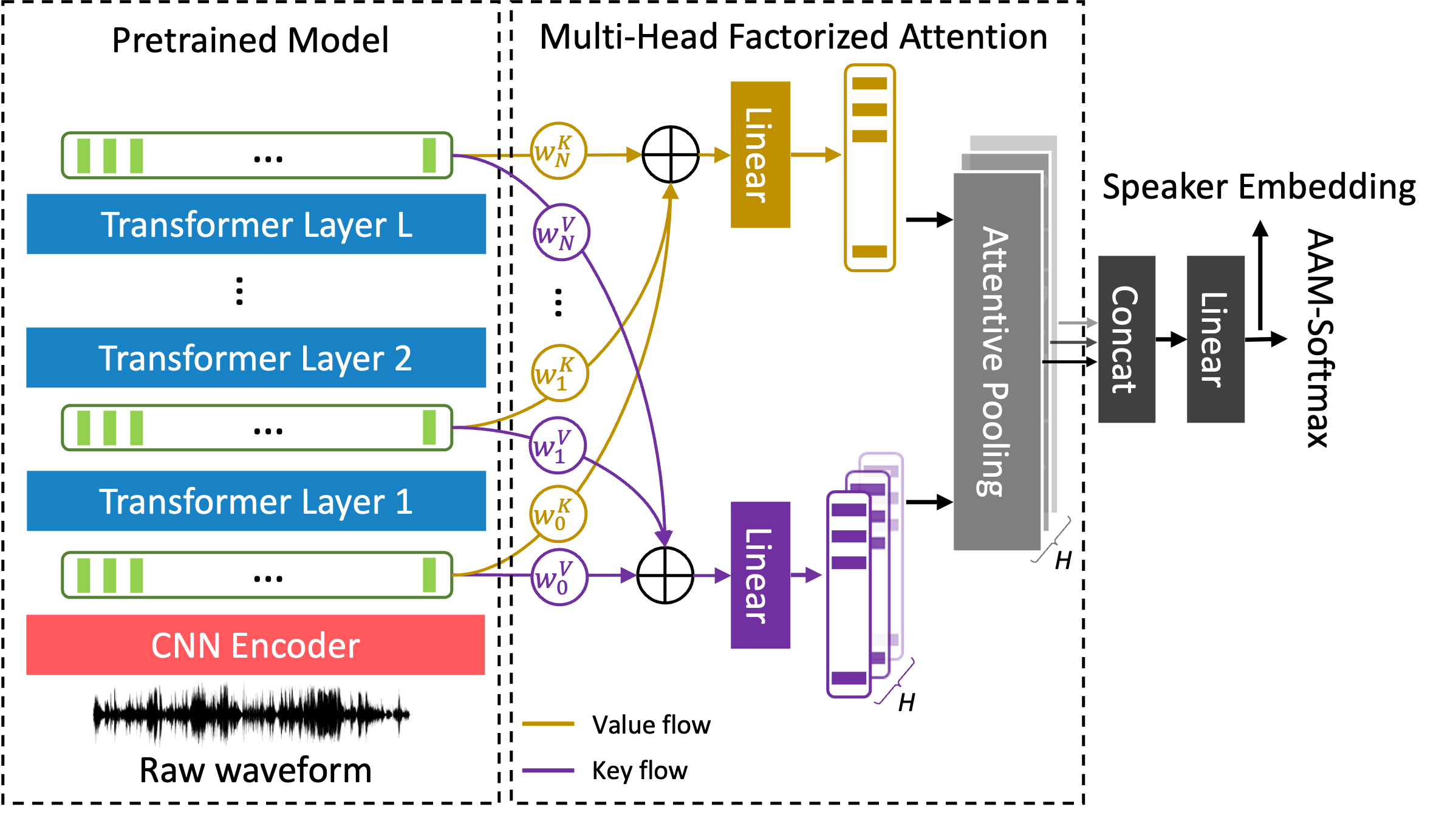}
    \caption{ (c) Proposed Multi-Head Factorized Attentive Pooling.}
    \label{fig:proposed}
\end{figure}

We study three light-weight back-ends attached to the pre-trained model for extracting speaker representation. All variants of the pre-trained model take raw waveform samples which are processed by a CNN encoder (see Figures \ref{fig:sys} and \ref{fig:proposed} and section \ref{sec:exp}). This part of the model is always fixed while the rest of the model, the Transformer encoder~\cite{vaswani2017attention}, is jointly optimized with the back-end during the fine-tuning.

\subsection{Top Layer Attentive Pooling}
\begin{figure*}[t]
\begin{minipage}[t]{.52\linewidth}
  \centering
  \centerline{\includegraphics[height=4.5cm]{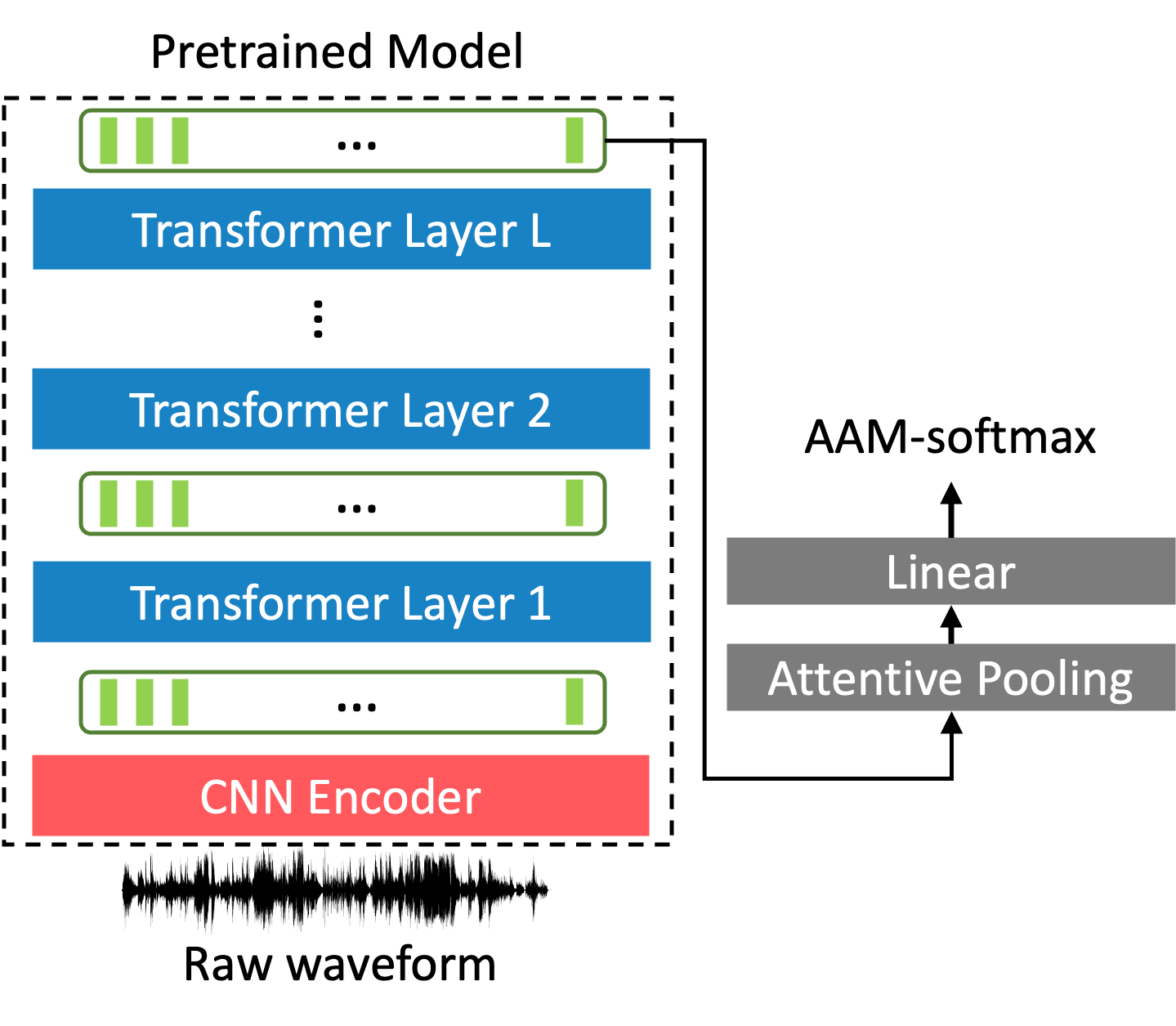}}
  \centerline{(a) Top Layer Attentive Pooling  }\medskip
\end{minipage}
\hfill
\begin{minipage}[t]{0.52\linewidth}
  \centering
  \centerline{\includegraphics[height=4.5cm]{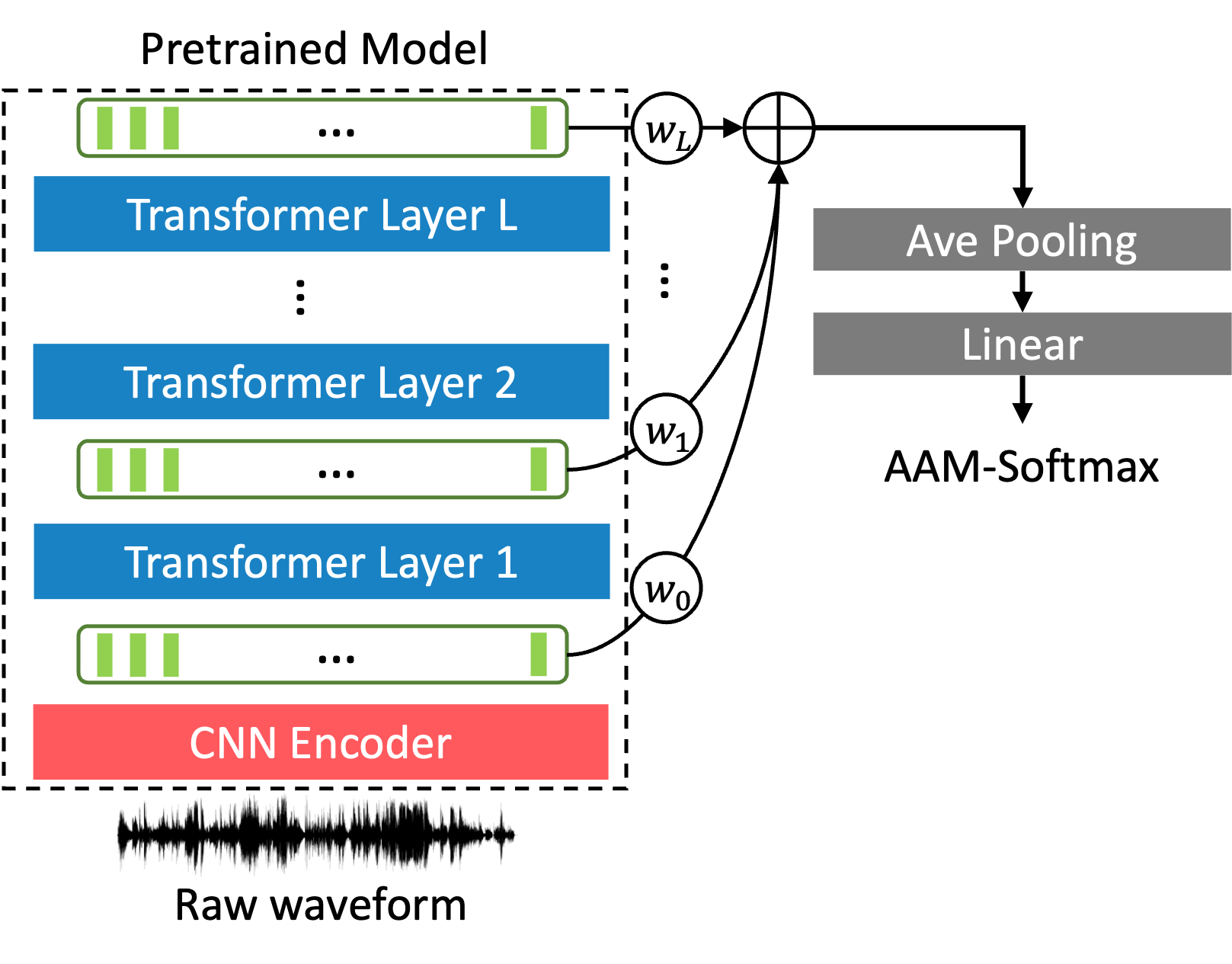}}
  \centerline{(b) Layer-wise Weighted Average Pooling }\medskip
\end{minipage} 
  
  \vfill
\caption{Architecture of the pre-trained model and attached light-weight back-end. During the fine-tuning, the CNN encoder is frozen, while the Transformer encoder and cascaded speaker extractor are jointly optimized.}
\label{fig:sys}
\end{figure*}

Intuitively, the pre-trained model can be regarded as a frame-level feature extractor that down-samples the raw waveform and then generates some universal frame-level features. Motivated by the standard x-vector style speaker extractor, we simply append an attentive statistical pooling layer and a linear layer on the top of the Transformer encoder as show in Fig \ref{fig:sys} (a). Compared to the standard x-vectors~\cite{snyder2018x}, the framewise TDNN part of the speaker embedding extractor is replaced by the pre-trained model.

\subsection{Layer-wise Weighted Average Pooling}
\label{subsec:LWWAP}

The large transformer models are trained on thousands of hours of speech data to well represent the structure of speech and obtain good generalization capability for various down-stream tasks. However, some studies using the last layer's representations \cite{fan2020exploring, wang2021fine, tak2022automatic} suggest that the advantage of fine-tuning pre-trained models over deep speaker extractors trained from scratch is insignificant or even non-existent. 
A potential reason is that the speech prediction objective of the pre-training stage is encouraging the model to ultimately discover and internally represent various acoustic units which in the last layers naturally correspond to (context-dependent) phones \cite{hsu2021hubert}. Therefore the top layers, which are closer to the objective of pre-training stage tend to be the most helpful for automatic speech recognition (ASR). In contrast, the speaker verification task benefits mainly from the low- and mid-level features, which carry most of the information about speaker identity. Thus, attempting to obtain the speaker representation only from the last Transformer layer's output might be sub-optimal for speaker verification.


As shown in Fig \ref{fig:sys} (b), following \cite{chen2021wavlm, chen2022unispeech}, we take the outputs of the $l$-th Transformer layer as $\mathbf{Z}_l \in \mathbb{R}^{T \times F}, l \in \{1,...,L\}$, where $L$ denotes the total number of Transformer layers in pre-trained model, $T$ is the number of total frames, and $F$ is the feature dimension of each frame. 
Then, each $\mathbf{Z}_l$ is assigned with a trainable weight $w_{l}$, and we learn a weighted average of all layers' outputs to generate frame-level features $\mathbf{O}\in \mathbb{R}^{T \times F}$ as follows:
\begin{equation}
\label{eq1}
\begin{split}
\mathbf{O} = \sum_{l=1}^{L} w_{l}\,\mathbf{Z}_{l}  
\end{split}
\end{equation}
Next, the weighted frame-level features are fed into an average pooling layer followed by a fully connected layer to obtain an utterance-level speaker embedding.

\subsection{Multi-Head Factorized Attentive Pooling}

In order to enhance the learned speaker representations, we employ an advanced attentive pooling, called multi-head factorized attentive pooling (MHFA), that clusters frame-level representations into acoustic units discovered by the transformer model. The frame-level representations are then aggregated (pooled) within each cluster and combined to produce the final speaker embedding. This mechanism allows speaker embeddings to be conditioned on the phonetic content of the input utterances. 

A similar approach has been successfully proposed for CNNs (e.g. in \cite{zhou2019cnn,li2019phonetic,stafylakis2019self}). Also note that our method shares similarities with i-vector methods that employ ASR to estimate the assignment of frames to speech recognition units, such as senones (e.g. in \cite{lei2014novel,kenny2014deep}). The crucial difference between our method and other phonetic-attention based approaches (e.g. \cite{zhou2019cnn}) is that ours is unsupervised (i.e. no transcribed data or pre-trained ASR model is required) and relies only on the capacity of pre-trained transformer-based architectures to capture meaningful acoustic/phonetic units (note that e.g. HuBERT acoustic units are highly correlated with phones \cite{hsu2021hubert}). 

To describe the attentive pooling, we use the \emph{query/key/value} abstraction for the attention mechanism as introduced in~\cite{vaswani2017attention}.
As mentioned above, we assume that the top layers of a pre-trained model contribute more to ASR task, while lower and middle layers contain more speaker information. Thus, we employ two sets of weights (factors) $\mathbf{w}^k$ and $\mathbf{w}^v$ to separately aggregate layer-wise outputs to produce the matrices of \emph{keys} $\mathbf{K}$ and \emph{values} $\mathbf{V}$, respectively:
\begin{equation}
\label{eq2}
\begin{split}
\mathbf{K} = \left(\sum_{l=1}^{L} w_{l}^k\mathbf{Z}_{l}\right)\mathbf{S}^k \\
\mathbf{V} = \left(\sum_{l=1}^{L} w_{l}^v\mathbf{Z}_{l}\right)\mathbf{S}^v, \\
\end{split}
\end{equation}
where matrices  $\mathbf{S}^k \in \mathbb{R}^{F \times D}$ and $\mathbf{S}^v \in \mathbb{R}^{F \times D}$ are used to reduce the dimensionality of keys and values, respectively.
Here, we assume that the model learns to produce values containing only speaker-discriminative information, while the keys might contain phonetic information, so that each attention head aggregates information from a specific set of phonetic units. This way the model can condition the speaker representation on and to be aware of the phonetic content of the input utterance. The architecture is shown in Fig~\ref{fig:proposed}.
To aggregate the values over frames, we use the following multi-head attention mechanism: 
\begin{equation}
\label{mhfa}
\begin{split}
\mathbf{A} = \mbox{softmax}\left(\mathbf{K}\mathbf{Q}\right), \\
\mathbf{c}_h = \sum_{t=1}^T \mathbf{A}_{h,t}  \mathbf{V}_t, \\
\mathbf{c} = \mbox{concat} \left(\mathbf{c}_1,...,\mathbf{c}_H\right), 
\end{split}
\end{equation}
where columns of matrix $\mathbf{Q} \in \mathbb{R}^{D \times H}$ correspond to learned query vectors of individual attention heads, $\mathbf{c}_h \in \mathbb{R}^{1 \times D} $ is the $h$-th sub-representation, and $\mathbf{c}\in\mathbb{R}^{1 \times HD}$ is the utterance-level speaker representation concatenated from all heads. Finally, the speaker label prediction is performed via passing the representation through a subsequent linear layer and $L_2$-normalization, and a classification layer that is needed only during training. In the testing stage, the $L_2$-normalized linear layer's output is regarded as the \emph{speaker embedding}.

\section{Methods for preventing over-fitting}
\label{sect:over-fit}
\subsection{Regularizing parameter during Fine-Tuning}
Although pre-trained models provide good initialization for speaker verification task, the fine-tuning process still tends to fall into local-optimum, especially on limited training data. This might be because the pre-trained models are much heavier than existing speaker extractor trained from scratch, and during the fine-tuning, such ``over-parameterized'' model will easily over-fit the small data and degenerate the performance.

To address this problem, we use L2 regularization towards the initial pre-trained model. This encourages the fine-tuned model's weights to remain close to the pre-trained initial ones, which leads to a more stable fine-tuning process. In the paper, we call it \emph{fine-tuning regularization}. The regularization term is computed as:

\begin{equation}
\label{eq3}
\begin{split}
\mathcal{L}_p = \sum_{j=1}^{\left|\Theta \right|} \left(\theta^j - \theta^j_{p}\right)^2, 
\end{split}
\end{equation}
where $\theta_{p} \in \Theta$ denotes the parameters inherited from the original pre-trained model, and $\theta \in \Theta$ denotes the current ones. Finally, the overall loss function $\mathcal{L}$ can be formulated as the combination of speaker classification loss $\mathcal{L}_{spk}$ and $\mathcal{L}_p$:
\begin{equation}
\label{eq4}
\begin{split}
\mathcal{L} = \mathcal{L}_{spk} + \lambda \mathcal{L}_p, 
\end{split}
\end{equation}
where $\lambda$ is a hyper-parameter to balance the two losses.

\subsection{Layer-wise Learning Rate Decay (LLRD) }
Intuitively, the bottom layers that are beneficial to the SV task should be assigned a lower learning rate to keep their function; meanwhile, the upper layers are supposed to be more flexible. In addition, each Transformer block is cascaded through a residual connection. If the shallow layers change rapidly, the original representation from the upper layers will completely change --  this might distort the modeling ability of the whole transformer encoder. 

Motivated by \cite{sun2019fine}, we adopt a layer-wise learning rate decay mechanism, assigning lower learning rates to the bottom layers and  higher ones to the top layers. Specifically, we first set a base learning rate $LR_{1}$ to the bottom layer, and then, an exponential decay rate is applied to the learning rates of each layer from bottom to top: 
\begin{equation}
\label{eq5}
\begin{split}
LR_{l} = LR_{1} \times \xi^{l-1} ,
\end{split}
\end{equation}
where $LR_{l}, l \in \{1,\dots,L\}$, denotes the learning rate of $l$-th Transformer layer, and $\xi$ is the weight decay factor. $\xi>1.0$ suggest higher learning rate for top layers, while $\xi =1$ indicates  that each layer is assigned with the same learning rate. The effect of $\xi$ will be further discussed in Sec~\ref{sec:rate-decay-exp}.

\begin{figure*}[t]
\begin{minipage}[t]{.32\linewidth}
  \centering
  \centerline{\includegraphics[width=5cm]{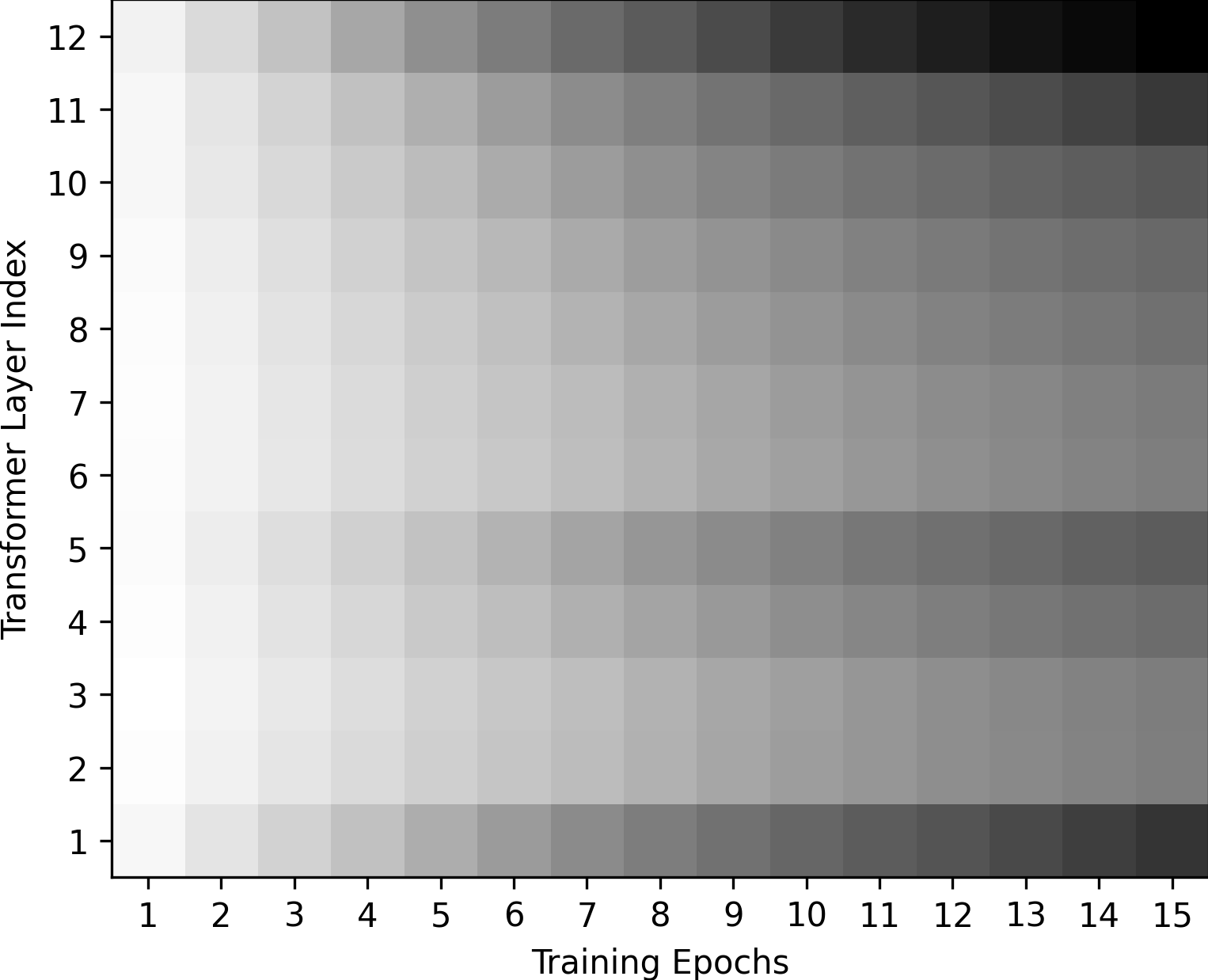}}
  \centerline{(a) Without constraints }\medskip
\end{minipage}
\hfill
\begin{minipage}[t]{0.32\linewidth}
  \centering
  \centerline{\includegraphics[width=5cm]{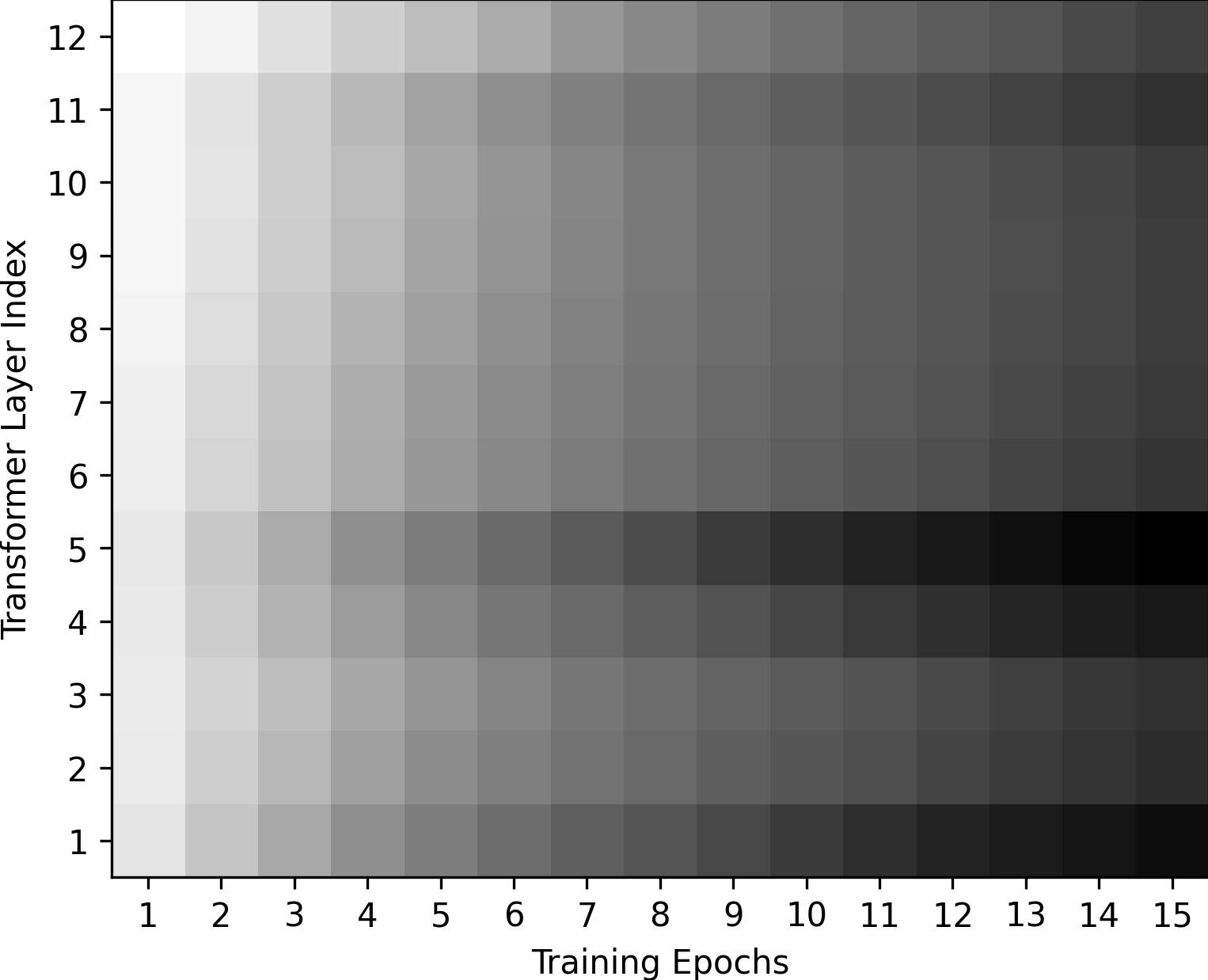}}
  \centerline{(b) Fine-tuning regularization}\medskip
\end{minipage}
  \hfill
\begin{minipage}[t]{.32\linewidth}
  \centering
  \centerline{\includegraphics[width=5cm]{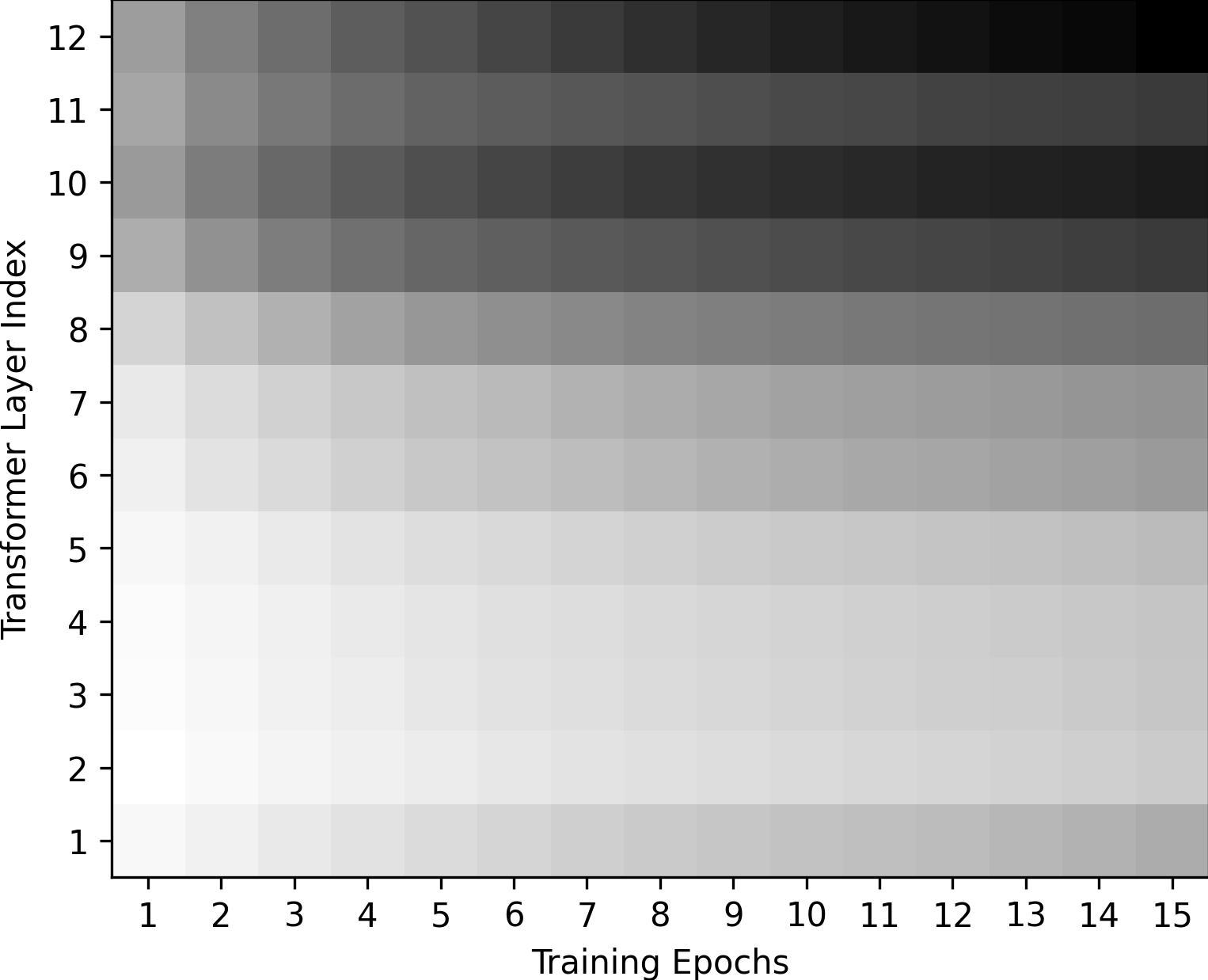}}
  \centerline{(c) Combining pre-trained reg and LLRD}\medskip
\end{minipage}
\caption{Parameter change trends. 
Brighter color reflects subtler change of optimized model compared to the initial pre-trained one. To better demonstrate the change trend, we use more training epochs.}
\label{fig:pt}

\end{figure*}

\section{Experiments}
\label{sec:exp}
\subsection{Setup}\textbf{Data-sets.} The SV performance is evaluated on the VoxCeleb corpus \cite{nagrani2017voxceleb,chung2018voxceleb2}, which is a widely used large-scale text-independent speaker verification data-set. The training set is derived from the development set of VoxCeleb2. The performance is evaluated on VoxCeleb1-O, VoxCeleb1-E, and VoxCeleb1-H trials.

\noindent\textbf{Implementation details.}
In this work, we utilize two types of pre-trained models: 1. The Base (Base+) model, including WavLM Base/Base+, HuBERT Base, Wav2Vec2.0 Base, contains a CNN encoder and 12 layers of Transformer. The dimension of the Transformer output is 768. The total number of parameters of those models are around 94M; 2. The Large model has 24 transformer blocks with 1024-dimensional output resulting in 316M parameters. All models are fine-tuned using 8 A100 GPUs with 10 epochs and are optimized by AAM-softmax \cite{deng2019arcface} with a margin of 0.2 and  scaling of 30. Hyper-parameter $\lambda$ is set to 1e-4. To speed up the training, the learning rate is decreased by 5\% each epoch. The duration of input raw waveform is set to 3 seconds. The mini-batch size of 120 is chosen for training all Base models, size 80 for the Large model. When adopting large margin fine-tuning (LM-FT) \cite{thienpondt2021idlab} to further boost performance, we input longer (5 seconds) waveforms and set the margin to 0.5 for additional 3 training epochs. Besides, Vox2-dev training data-set is augmented by adding noise and reverberation (MUSAN and RIR data-sets) \cite{chung2020in}.  

\noindent\textbf{Performance Metrics.}
Both equal error rate (EER) and minimum detection cost function (minDCF) are employed to measure the performances of speaker verification systems. The target probability $P_{\textit{tar}}$ is set to 0.01 or 0.05, for DCF1 and DCF5, respectively. $C_{\textit{fa}}$ and $C_{\textit{fr}}$ share the equal weight of 1.0.

\subsection{Results}
\label{sec:rlt}
\subsubsection{Speaker Extractor Variants}
The performance of our speaker extractor variants is reported in Table \ref{tab:1}. Note that for (c) $H$ is set to 8.
\begin{table}[tbh] 
    \centering
        \caption{Comparison of pooling methods shown in Figs. \ref{fig:sys}, \ref{fig:proposed},  on VoxCeleb1-O test data-set using WavLM Base+.}
    \begin{tabular}{l|c|c}
    \hline
        Variants & EER(\%) & DCF1 \\
        \hline
        \hline

        (a) Layer-Attentive Pooling & 1.00 & 0.148 \\
        (b) Layer-wise Weighted Average Pooling & 1.16 & 0.113 \\
        (c) Multi-Head Factorized Attention & 0.92 & 0.112 \\
        \hline
        
    \end{tabular}
    \label{tab:1}
\end{table}

\begin{figure}[htb]
  \centering
  \includegraphics[width=0.8\linewidth]{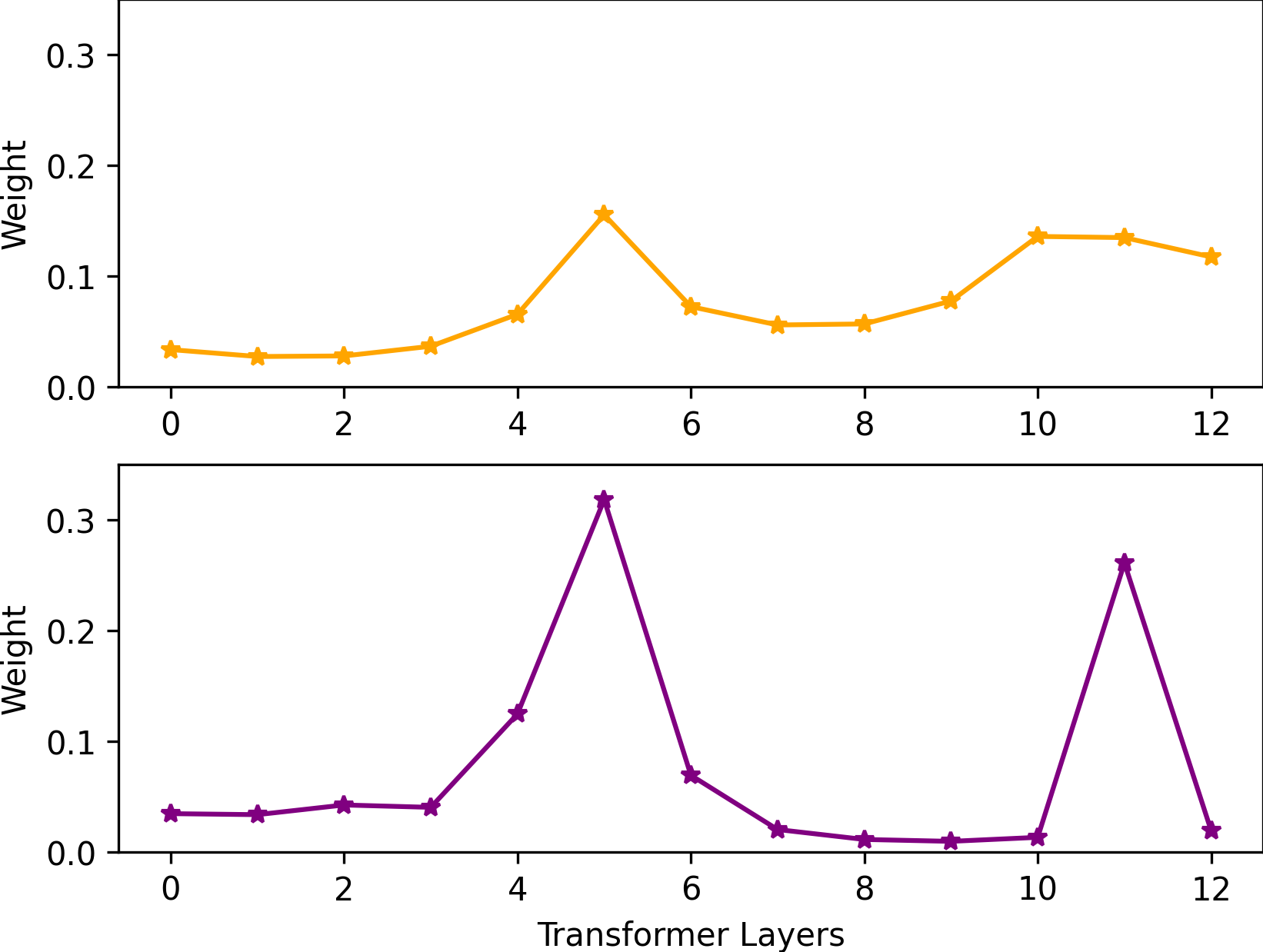}
\caption{Layer-wise weights of value flow (yellow) and key flow (purple). 
$0$-th Transformer layer denotes the output of the CNN encoder, which is also the input of the 1st Transformer layer.}
\label{fig:kv}
\end{figure}
The MHFA based system (c) outperforms layer-wise weighted average pooling based system (b). This suggests that using multi-factor to model speaker information and potential phonetic information is able to yield more discriminative speaker representations. Moreover, we also complement our experiments by analyzing the weights of key $w^k$ and value $w^v$ in Fig~\ref{fig:kv}. Compared to $w^v$, $w^k$ gives a lot of attention to the $11$-th Transformer's output, which is highly consistent with the conclusion reported in \cite{chen2021wavlm} that $11$-th layer contributes significantly to phoneme recognition task. This finding justifies the benefits of the proposed multi-head factorized attentive pooling exploiting the acoustic units derived from HuBERT or WavLM. They enable the model to assign frames of specific phonetic content to the corresponding attention head, independently of speaker or channel characteristics. 


\subsubsection{Different constraints of fine-tuning}
We experiment with various constraints to stabilize the fine-tuning in Table \ref{tab:Limitions}. We observed that restriction on the distributions of $w^k$ and $w^v$, i.e. reinforcing their similarity, harms the performance. 
In addition, sharing a linear layer (i.e. $\mathbf{S}^k=\mathbf{S}^v$) leads to poor performance. This suggests that it is difficult to separate speaker information and phoneme information simultaneously with a single linear layer, which further confirms the effectiveness of the proposed factorization. 
The best EER result of 0.85\%  is achieved by applying fine-tuning regularization. This indicates that an explicit bias towards the initial model is beneficial for the discriminability of the speaker representations and for avoiding over-fitting. 
\begin{table}[tb] 
    \centering
        \caption{Results using WavLM Base+ on Vox1-O using different constraints. KV represents the constraint between Key and Value weights. }
    \begin{tabular}{l|c|c}
    \hline
        Constraints & EER(\%) & DCF1 \\
        \hline
        \hline
    %
        No sharing between KV & 0.92 & 0.112\\
        KV Shared Layer-wise Weights & 0.96 & 0.122 \\
        KV Shared Linear Layer & 1.01 & 0.110 \\
        KV Shared Linear Layer + Weights & 0.99 & 0.123 \\
        Fine-tuning regularization & 0.85 & 0.113 \\
        \hline
        
    \end{tabular}                                       
    \label{tab:Limitions}
\end{table}
\subsubsection{Layer-wise learning rate decay}
\label{sec:rate-decay-exp}
Table \ref{tab:LLRD} presents the results of using different learning rate schedules. The base learning rate of MHFA is set to 1e-3 for all experiments. As expected, freezing the parameters of the pretrained model leads to the worst performance. Using the same high learning rate (i.e. 1e-3) for both training the MHFA back-end and fine-tuning the Transformer encoder in the pre-trained model did not lead to convergence. However, when training the the MHFA back-end (from random initialization) with learning rate 1e-3 and simultaneously fine-tuning the Transformer encoder with a low learning rate an excellent performance can be achieved. In addition, when $\xi > 1$, which means the learning rate increases from bottom to top, the best performance is achieved (EER: 0.80\%). Note that our training procedure does not require the back-end to be first trained using the fixed pre-trained model, which leads to substantial computational savings during the training.
\begin{table}[tb]
    \centering
        \caption{Comparison of different layer-wise rates. Weight decay factor $\xi <1$ means lower learning rates for top layers. For fair comparison, all systems utilize fine-tuning regularization. }
    \begin{tabular}{c|c|c|c|c}
    \hline
    \multicolumn{1}{c|}{\multirow{1}{*}{MHFA}}&
    \multicolumn{2}{c|}{Transformer Encoder}& \multirow{2}{*}{EER(\%)} & \multirow{2}{*}{DCF1}
    
     \cr\cline{1-3} LR & LR & $\xi$ &&  \cr 
        \hline
        \hline
        1e-3 & 1e-3 & 1.0 & NAN & NAN \\
        \hline
        1e-3 & Fixed & - & 2.16 & 0.256 \\
        \hline
        \multirow{6}{*}{1e-3} & \multirow{6}{*}{2e-5} & 0.6 & 1.56 & 0.182 \\
        &&0.8 & 1.49 & 0.174 \\
        &&1.0 & 0.85 & 0.113 \\
        &&1.5 & \textbf{0.80} & \textbf{0.088} \\
        &&1.8 & 0.84 & 0.114 \\
        &&2.0 & 2.68 & 0.309 \\
        \hline
        
    \end{tabular}
    \label{tab:LLRD}

\end{table}

To further understand how and why LLRD works, we track the parameter trends of each Transformer block with different settings. As shown in Fig~\ref{fig:pt}, compared to unrestricted fine-tuning (a), applying fine-tuning regularization (b) makes the change of each layer's parameters smoother instead of drastically optimizing a few layers. When combining the fine-tuning regularization and LLRD (c), we find that the parameter changes of lower layers tend to be smaller. This observation meets our expectation that bottom layers are more sensitive to speaker-related task.

\begin{table*}[h]
  \caption{Results for speaker verification on the Voxceleb1-O data-set and extended VoxCeleb1-E and VoxCeleb-H test sets. All models are trained on VoxCeleb2-dev, except $\dag$ -- its training data consists of VoxCeleb2 and VoxCeleb-dev. LM-FT denotes large margin fine-tuning. }
  \label{tab:vsSOTA}
  \centering
    \begin{tabular}{c|c|c|c|c|c|c|c|c|c|c}  
    \hline  
    \multicolumn{1}{c|}{\multirow{2}{*}{Front-end Model}}&\multicolumn{1}{c|}{\multirow{2}{*}{Params}}&\multicolumn{3}{c|}{VoxCeleb1-O}&\multicolumn{3}{c}{VoxCeleb1-E}&\multicolumn{3}{|c}{VoxCeleb1-H}\cr\cline{3-11}&&EER&DCF1&DCF5&EER&DCF1&DCF5&EER&DCF1&DCF5\cr 
    \hline
    \hline
     ECAPA-TDNN \cite{kwon2021ins} & 14.7M &  0.90 & - & 0.081 & 1.11 & - & 0.077 & 2.32 & - & 0.155 \\
    
      
    
    TSE-Spine2Net \cite{rybicka2021spine2net} & 58.0M & 0.92 & 0.105 & 0.068 & 0.99 & 0.112 & 0.065 & 1.95 & 0.192 & 0.117 \\

    Wav2Vec-TDNN $^{\dag}$ \cite{novoselov2022robust} & 317M + 3M & 0.84 & 0.058 & - & -& - &- &- & - & -\\
    UnispeechSAT\_Base-TDNN \cite{chen2022large} & 94M + 6M &  1.00 & - & - & 0.93 & - & - & 1.87 & - & - \\
    
    WavLM\_Base+-TDNN \cite{chen2021wavlm} & 94M + 6M  & 0.84 & - & - & 0.92 & - & - & 1.75 & - & - \\
    
    HuBERT\_Large-TDNN \cite{chen2022large} & 316M + 6M & 0.59 & - & - & 0.65 & - & - & 1.22 & - & -  \\
    UnispeechSAT\_Large-TDNN \cite{chen2022large} & 316M + 6M  & 0.53 & - & - & 0.56 & - & - & 1.18 & - & - \\
    
    WavLM\_Large-TDNN \cite{chen2021wavlm} & 316M + 6M  & 0.38 & - & - & 0.48 & - & - & 0.98 & - & - \\
    \hline
    \hline
    \multicolumn{11}{l}{pre-trained Model: \textbf{Wav2Vec 2.0 Base}, pre-training corpusrpus: 960 hr} \\
    \hline
    \multicolumn{1}{l|}{without LLRD and fine-tuning reg} & 95M + 0.4M  & 2.31 & 0.266 & 0.164 & 2.24 & 0.273 & 0.159 & 4.97 & 0.464 & 0.306 \\
    \multicolumn{1}{l|}{\textbf{Proposed}} & 95M + 0.4M  & 1.92 & 0.229 & 0.138 & 2.04 & 0.243 & 0.144 & 4.44 & 0.421 & 0.277 \\
    \multicolumn{1}{l|}{\textbf{Proposed+LM-FT}} & 95M + 0.4M  & 1.75 & 0.204 & 0.127 & 1.93 & 0.224 & 0.131 & 4.10 & 0.401 & 0.262 \\
    \hline
    \multicolumn{11}{l}{pre-trained Model: \textbf{HuBERT Base}, pre-training corpus: 960 hr} \\
    \hline
    \multicolumn{1}{l|}{without LLRD and fine-tuning reg} & 94M + 0.4M  & 1.13 & 0.157 & 0.089 & 1.42 & 0.160 & 0.092 & 2.98 & 0.297 & 0.185 \\
    \multicolumn{1}{l|}{\textbf{Proposed}} & 94M + 0.4M  & 0.97 & 0.110 & 0.067 & 1.28 & 0.141 & 0.083 & 2.65 & 0.263 & 0.162 \\
    \multicolumn{1}{l|}{\textbf{Proposed+LM-FT}} & 94M + 0.4M  & 0.92 & 0.091 & 0.059 & 1.19 & 0.136 & 0.078 & 2.52 & 0.252 & 0.159 \\
    \hline
    \multicolumn{11}{l}{pre-trained Model: \textbf{WavLM Base}, pre-training corpus: 960 hr} \\
    \hline
    \multicolumn{1}{l|}{without LLRD and fine-tuning reg} & 94M + 0.4M  & 1.17 & 0.131 & 0.078 & 1.36 & 0.154 & 0.087 & 2.80 & 0.270 & 0.172 \\
    \multicolumn{1}{l|}{\textbf{Proposed}} & 94M + 0.4M  & 0.96 & 0.110 & 0.068 & 1.19 & 0.131 & 0.078 & 2.49 & 0.250 & 0.156 \\
    \multicolumn{1}{l|}{\textbf{Proposed+LM-FT}} & 94M + 0.4M  & 0.89 & 0.094 & 0.056 & 1.09 & 0.117 & 0.068 & 2.27 & 0.231 & 0.142 \\
    \hline
    \multicolumn{11}{l}{pre-trained Model: \textbf{WavLM Base+}, pre-training corpus: 94,000 hr } \\
    \hline
    \multicolumn{1}{l|}{without LLRD and fine-tuning reg} & 94M + 0.4M  & 0.92 & 0.112 & 0.063 & 1.18 & 0.154 & 0.081 & 2.18 & 0.218 & 0.135 \\
    \multicolumn{1}{l|}{\textbf{Proposed [8 heads]}} & 94M + 0.4M  & 0.80 & 0.088 & 0.055 & 1.07 & 0.121 & 0.070 & 2.20 & 0.222 & 0.137 \\
    \multicolumn{1}{l|}{\textbf{Proposed [8 heads] + LM-FT}} & 94M + 0.4M  & 0.76 & 0.074 & 0.050 & 0.95 & 0.109 & 0.063 & 2.01 & 0.208 & 0.124 \\ 
    \hline
    \multicolumn{1}{l|}{\textbf{Proposed [16 heads]}} & 94M + 0.7M  & 0.77 & 0.081 & 0.055 & 0.95 & 0.108 & 0.061 & 2.03 & 0.203 & 0.128 \\
    \multicolumn{1}{l|}{\textbf{Proposed [32 heads]}} & 94M + 1.2M  & 0.71 & 0.078 & 0.048 & 0.90 & 0.100 & 0.059 & 1.97 & 0.198 & 0.122  \\
    \multicolumn{1}{l|}{\textbf{Proposed [64 heads]}} & 94M + 2.2M  & 0.66 & 0.074 & 0.045 & 0.89 & 0.097 & 0.056 & 1.90 & 0.190 & 0.119 \\
    \multicolumn{1}{l|}{\textbf{Proposed [128 heads]}} & 94M + 4.3M  & 0.72 & 0.069 & 0.045 & 0.87 & 0.097 & 0.056 & 1.82 & 0.191 & 0.113  \\
    \multicolumn{1}{l|}{\textbf{Proposed [64 heads] + LM-FT}} & 94M + 2.2M  & \textbf{0.59} & \textbf{0.069} & \textbf{0.041} & \textbf{0.79} & \textbf{0.089} & \textbf{0.050} & \textbf{1.73} & \textbf{0.177} & \textbf{0.107} \\ 
    \hline
    \multicolumn{11}{l}{pre-trained Model: \textbf{WavLM Large}, pre-training corpus: 94,000 hr} \\
    \hline
    \multicolumn{1}{l|}{without LLRD and fine-tuning reg} & 316M + 2.2M  & 1.02 & 0.134 & 0.081 & 1.02 & 0.118 & 0.067 & 2.29 & 0.230 & 0.145 \\
    \multicolumn{1}{l|}{\textbf{Proposed [64 heads]}} & 316M + 2.2M  & 0.49 & 0.081 & 0.041 & 0.70 & 0.091 & 0.051 & 1.70 & 0.177 & 0.105 \\
    \multicolumn{1}{l|}{\textbf{Proposed [64 heads] + LM-FT}} & 316M + 2.2M  & 0.49 & 0.091 & 0.045 & 0.80 & 0.084 & 0.049 & 1.70 & 0.163 & 0.101 \\

    
    \hline
    \hline
    \end{tabular}
\end{table*}
\subsubsection{Comparison with state-of-the-art systems}
We compare the proposed method with other state-of-the-art systems using pre-trained models in Table \ref{tab:vsSOTA}. The results show a significant degradation in performance when the LLRD and fine-tuning regularization is not used during the fine-tuning. This is especially true for the Large pre-trained model (i.e. WavLM Large). We observe that with the number of heads $H$ increasing (i.e. from 8 to 128), the system performance improves and with $H=64$, the performance starts to peak. 
Further improvement can be obtained by adding the large margin fine-tuning strategy, which leads to very competitive results (the bold numbers in Table~\ref{tab:vsSOTA}) as compared to the SOTA methods utilizing a pre-trained model of a similar size (94M parameters).
We also include our preliminary experiments with WavLM Large (the last block in Table~\ref{tab:vsSOTA}), where we achieved similar or better results as with the Base+ model, but we did not attempt to extensively adjust our fine-tuning strategy. 

\subsubsection{Comparison of pre-trained models}
From Table \ref{tab:vsSOTA}, our results show a large gap in performance between Wav2Vec 2.0 and HuBERT, indicating that the training paradigm of the latter encourages the network to model speaker information in the intermediate layers. One plausible explanation is that the iterative clustering mechanism of HuBERT (i.e. k-means representations from an intermediate Transformer layer, prediction of masked tokens' cluster index in the last layer) yields more speaker-independent targets, and therefore encourages the network to model speaker information in order to use it for speaker normalization in the final layers (i.e. to remove speaker information in the output). This is in contrast to the Wav2Vec 2.0 training approach, where (a) the discretization is taking place on the outputs of feature encoder (i.e. the CNN), i.e. on representations that clearly contain speaker information, and (b) the classification task is between the true cluster index and those of distractors, which are uniformly sampled from other masked time steps of the same utterance (i.e. contrastive loss).

The performance of WavLM is only marginally better than that of HuBERT, despite the fact that the training scheme is more targeted to speaker modeling (i.e. use of simulated noisy/overlapped speech as inputs, and prediction of the pseudo-labels of original speech on the masked region \cite{chen2021wavlm}). Nevertheless, this training scheme, combined with the architectural improvements (gated relative position bias \cite{chen2021wavlm}) yields some improvements over HuBERT.

\section{Conclusion}
We proposed an attention-based back-end and a fine-tuning strategy for attaining state-of-the-art results in speaker recognition using pre-trained  self-supervised speech models with transformer architectures. The back-end utilizes all transformer layers and creates two feature streams via two learnable layer-wise aggregators. The \emph{key} stream is used for estimating the phonetic content of the utterance while the \emph{value} stream carries the speaker-discriminative information. The proposed method is tested with five different self-supervised models, and the comparisons with other pooling strategies and published results demonstrates its effectiveness. Furthermore, our method is the first to attain such state-of-the-art results without frame-level convolutional layers, indicating the power of Transformers and attention in extracting speaker information. Finally, we proposed a fine-tuning strategy combining layer-wise learning rate decay and a regularization method that encourages the model parameters to remain close to the ones of the pre-trained model. This strategy was crucial to reach state-of-the-art performance and can be further explored, e.g. by using the Fisher Information Matrix in the regularizer, as proposed in Elastic Weight Consolidation \cite{kirkpatrick2017overcoming}.    
\label{sec:conclusion}

\section{Acknowledgements}
The work was partly supported by Czech National Science Foundation (GACR) project NEUREM3 No. 19-26934X, Czech Ministry of Education, Youth and Sports from project no. LTAIN19087 "Multi-linguality in speech technologies", and Horizon 2020 Marie Sklodowska-Curie grant ESPERANTO, No. 101007666. Computing on IT4I supercomputer was supported by the Czech Ministry of Education, Youth and Sports from the Large Infrastructures for Research, Experimental Development and Innovations project "e-Infrastructure CZ – LM2018140".

\bibliographystyle{IEEEbib}
\bibliography{mybib}

\begin{thebibliography}{10}

\bibitem{snyder2018x}
David Snyder, Daniel Garcia-Romero, Gregory Sell, Daniel Povey, and Sanjeev
  Khudanpur,
\newblock ``{X-vectors: Robust DNN embeddings for speaker recognition},''
\newblock in {\em 2018 IEEE International Conference on Acoustics, Speech and
  Signal Processing (ICASSP)}. IEEE, 2018, pp. 5329--5333.

\bibitem{desplanques2020ecapa}
Brecht Desplanques, Jenthe Thienpondt, and Kris Demuynck,
\newblock ``{ECAPA-TDNN: Emphasized Channel Attention, Propagation and
  Aggregation in TDNN Based Speaker Verification},''
\newblock in {\em Interspeech2020}, 2020, pp. 1--5.

\bibitem{schneider2019Wav2Vec}
Steffen Schneider, Alexei Baevski, Ronan Collobert, and Michael Auli,
\newblock ``{wav2vec: Unsupervised Pre-Training for Speech Recognition},''
\newblock {\em Proc. Interspeech 2019}, pp. 3465--3469, 2019.

\bibitem{baevski2020Wav2Vec}
Alexei Baevski, Yuhao Zhou, Abdelrahman Mohamed, and Michael Auli,
\newblock ``wav2vec 2.0: A framework for self-supervised learning of speech
  representations,''
\newblock {\em Advances in Neural Information Processing Systems}, vol. 33, pp.
  12449--12460, 2020.

\bibitem{hsu2021hubert}
Wei-Ning Hsu, Benjamin Bolte, Yao-Hung~Hubert Tsai, Kushal Lakhotia, Ruslan
  Salakhutdinov, and Abdelrahman Mohamed,
\newblock ``{HuBERT: Self-Supervised Speech Representation Learning by Masked
  Prediction of Hidden Units},''
\newblock {\em IEEE/ACM Transactions on Audio, Speech, and Language
  Processing}, vol. 29, pp. 3451--3460, 2021.

\bibitem{chen2021wavlm}
Sanyuan Chen, Chengyi Wang, Zhengyang Chen, Yu~Wu, Shujie Liu, Zhuo Chen, Jinyu
  Li, Naoyuki Kanda, Takuya Yoshioka, Xiong Xiao, Jian Wu, Long Zhou, Shuo Ren,
  Yanmin Qian, Yao Qian, Jian Wu, Michael Zeng, Xiangzhan Yu, and Furu Wei,
\newblock ``{WavLM: Large-Scale Self-Supervised Pre-Training for Full Stack
  Speech Processing},''
\newblock {\em IEEE Journal of Selected Topics in Signal Processing}, pp.
  1--14, 2022.

\bibitem{chen2022unispeech}
Sanyuan Chen, Yu~Wu, Chengyi Wang, Zhengyang Chen, Zhuo Chen, Shujie Liu, Jian
  Wu, Yao Qian, Furu Wei, Jinyu Li, et~al.,
\newblock ``{Unispeech-Sat: Universal Speech Representation Learning With
  Speaker Aware Pre-Training},''
\newblock in {\em ICASSP 2022-2022 IEEE International Conference on Acoustics,
  Speech and Signal Processing (ICASSP)}. IEEE, 2022, pp. 6152--6156.

\bibitem{yang2021superb}
Shu-wen Yang, Po-Han Chi, Yung-Sung Chuang, Cheng-I~Jeff Lai, Kushal Lakhotia,
  Yist~Y Lin, Andy~T Liu, Jiatong Shi, Xuankai Chang, Guan-Ting Lin, et~al.,
\newblock ``Superb: Speech processing universal performance benchmark,''
\newblock {\em arXiv preprint arXiv:2105.01051}, 2021.

\bibitem{fan2020exploring}
Zhiyun Fan, Meng Li, Shiyu Zhou, and Bo~Xu,
\newblock ``Exploring wav2vec 2.0 on speaker verification and language
  identification,''
\newblock {\em arXiv preprint arXiv:2012.06185}, 2020.

\bibitem{novoselov2022robust}
Sergey Novoselov, Galina Lavrentyeva, Anastasia Avdeeva, Vladimir Volokhov, and
  Aleksei Gusev,
\newblock ``Robust speaker recognition with transformers using wav2vec 2.0,''
\newblock {\em arXiv preprint arXiv:2203.15095}, 2022.

\bibitem{chen2022large}
Zhengyang Chen, Sanyuan Chen, Yu~Wu, Yao Qian, Chengyi Wang, Shujie Liu, Yanmin
  Qian, and Michael Zeng,
\newblock ``{Large-Scale Self-Supervised Speech Representation Learning for
  Automatic Speaker Verification},''
\newblock in {\em ICASSP 2022-2022 IEEE International Conference on Acoustics,
  Speech and Signal Processing (ICASSP)}. IEEE, 2022, pp. 6147--6151.

\bibitem{yang21c_interspeech}
Shu wen Yang, Po-Han Chi, Yung-Sung Chuang, Cheng-I~Jeff Lai, Kushal Lakhotia,
  Yist~Y. Lin, Andy~T. Liu, Jiatong Shi, Xuankai Chang, Guan-Ting Lin,
  Tzu-Hsien Huang, Wei-Cheng Tseng, Ko~tik Lee, Da-Rong Liu, Zili Huang, Shuyan
  Dong, Shang-Wen Li, Shinji Watanabe, Abdelrahman Mohamed, and Hung yi~Lee,
\newblock ``{SUPERB: Speech Processing Universal PERformance Benchmark},''
\newblock in {\em Proc. Interspeech 2021}, 2021, pp. 1194--1198.

\bibitem{wang2021fine}
Yingzhi Wang, Abdelmoumene Boumadane, and Abdelwahab Heba,
\newblock ``A fine-tuned wav2vec 2.0/hubert benchmark for speech emotion
  recognition, speaker verification and spoken language understanding,''
\newblock {\em arXiv preprint arXiv:2111.02735}, 2021.

\bibitem{vaessen2022fine}
Nik Vaessen and David~A Van~Leeuwen,
\newblock ``Fine-tuning wav2vec2 for speaker recognition,''
\newblock in {\em ICASSP 2022-2022 IEEE International Conference on Acoustics,
  Speech and Signal Processing (ICASSP)}. IEEE, 2022, pp. 7967--7971.

\bibitem{vaswani2017attention}
Ashish Vaswani, Noam Shazeer, Niki Parmar, Jakob Uszkoreit, Llion Jones,
  Aidan~N Gomez, {\L}ukasz Kaiser, and Illia Polosukhin,
\newblock ``Attention is all you need,''
\newblock {\em Advances in neural information processing systems}, vol. 30,
  2017.

\bibitem{tak2022automatic}
Hemlata Tak, Massimiliano Todisco, Xin Wang, Jee-weon Jung, Junichi Yamagishi,
  and Nicholas Evans,
\newblock ``Automatic speaker verification spoofing and deepfake detection
  using wav2vec 2.0 and data augmentation,''
\newblock {\em arXiv preprint arXiv:2202.12233}, 2022.

\bibitem{zhou2019cnn}
Tianyan Zhou, Yong Zhao, Jinyu Li, Yifan Gong, and Jian Wu,
\newblock ``{CNN with phonetic attention for text-independent speaker
  verification},''
\newblock in {\em 2019 IEEE Automatic Speech Recognition and Understanding
  Workshop (ASRU)}. IEEE, 2019, pp. 718--725.

\bibitem{li2019phonetic}
Lantian Li, Zhiyuan Tang, Ying Shi, and Dong Wang,
\newblock ``Phonetic-attention scoring for deep speaker features in speaker
  verification,''
\newblock in {\em 2019 Asia-Pacific Signal and Information Processing
  Association Annual Summit and Conference (APSIPA ASC)}. IEEE, 2019, pp.
  284--288.

\bibitem{stafylakis2019self}
Themos Stafylakis, Johan Rohdin, Old{\v{r}}ich Plchot, Petr Mizera, and
  Luk{\'a}{\v{s}} Burget,
\newblock ``Self-supervised speaker embeddings,''
\newblock {\em Proc. Interspeech 2019}, pp. 2863--2867, 2019.

\bibitem{lei2014novel}
Yun Lei, Nicolas Scheffer, Luciana Ferrer, and Mitchell McLaren,
\newblock ``A novel scheme for speaker recognition using a phonetically-aware
  deep neural network,''
\newblock in {\em 2014 IEEE international conference on acoustics, speech and
  signal processing (ICASSP)}. IEEE, 2014, pp. 1695--1699.

\bibitem{kenny2014deep}
Patrick Kenny, Themos Stafylakis, Pierre Ouellet, Vishua Gupta, and Jahangir
  Md~Alam,
\newblock ``Deep neural networks for extracting baum-welch statistics for
  speaker recognition.,''
\newblock in {\em Odyssey 2014: The Speaker and Language Recognition Workshop},
  2014.

\bibitem{sun2019fine}
Chi Sun, Xipeng Qiu, Yige Xu, and Xuanjing Huang,
\newblock ``How to fine-tune bert for text classification?,''
\newblock in {\em China national conference on Chinese computational
  linguistics}. Springer, 2019, pp. 194--206.

\bibitem{nagrani2017voxceleb}
Arsha Nagrani, Joon~Son Chung, and Andrew Zisserman,
\newblock ``Voxceleb: A large-scale speaker identification dataset,''
\newblock {\em Proc. Interspeech 2017}, pp. 2616--2620, 2017.

\bibitem{chung2018voxceleb2}
Joon~Son Chung, Arsha Nagrani, and Andrew Zisserman,
\newblock ``Voxceleb2: Deep speaker recognition,''
\newblock {\em Proc. Interspeech 2018}, pp. 1086--1090, 2018.

\bibitem{deng2019arcface}
Jiankang Deng, Jia Guo, Niannan Xue, and Stefanos Zafeiriou,
\newblock ``Arcface: Additive angular margin loss for deep face recognition,''
\newblock in {\em Proceedings of the IEEE/CVF conference on computer vision and
  pattern recognition}, 2019, pp. 4690--4699.

\bibitem{thienpondt2021idlab}
Jenthe Thienpondt, Brecht Desplanques, and Kris Demuynck,
\newblock ``The idlab voxsrc-20 submission: Large margin fine-tuning and
  quality-aware score calibration in dnn based speaker verification,''
\newblock in {\em ICASSP 2021-2021 IEEE International Conference on Acoustics,
  Speech and Signal Processing (ICASSP)}. IEEE, 2021, pp. 5814--5818.

\bibitem{chung2020in}
Joon~Son Chung, Jaesung Huh, Seongkyu Mun, Minjae Lee, Hee~Soo Heo, Soyeon
  Choe, Chiheon Ham, Sunghwan Jung, Bong-Jin Lee, and Icksang Han,
\newblock ``In defence of metric learning for speaker recognition,''
\newblock in {\em Proc. Interspeech}, 2020.

\bibitem{kwon2021ins}
Yoohwan Kwon, Hee-Soo Heo, Bong-Jin Lee, and Joon~Son Chung,
\newblock ``{The ins and outs of speaker recognition: lessons from VoxSRC
  2020},''
\newblock in {\em ICASSP 2021-2021 IEEE International Conference on Acoustics,
  Speech and Signal Processing (ICASSP)}. IEEE, 2021, pp. 5809--5813.

\bibitem{rybicka2021spine2net}
Magdalena Rybicka, Jes{\'u}s Villalba, Piotr Zelasko, Najim Dehak, and Konrad
  Kowalczyk,
\newblock ``{Spine2Net: SpineNet with Res2Net and Time-Squeeze-and-Excitation
  Blocks for Speaker Recognition},''
\newblock {\em Proc. Interspeech 2021}, pp. 496--500, 2021.

\bibitem{kirkpatrick2017overcoming}
James Kirkpatrick, Razvan Pascanu, Neil Rabinowitz, Joel Veness, Guillaume
  Desjardins, Andrei~A Rusu, Kieran Milan, John Quan, Tiago Ramalho, Agnieszka
  Grabska-Barwinska, et~al.,
\newblock ``Overcoming catastrophic forgetting in neural networks,''
\newblock {\em Proceedings of the national academy of sciences}, vol. 114, no.
  13, pp. 3521--3526, 2017.

\end{thebibliography}

\end{document}